\documentclass[usenatbib,useAMS]{mnras}
\usepackage{txfonts,amssymb}
\usepackage{amsfonts}
\usepackage{epsfig}

\title[New counterjet in NGC~1275]{Discovery of a new subparsec
 counterjet in NGC~1275: the inclination angle and the environment}

\author[Y. Fujita and H. Nagai]{Yutaka Fujita,$^{1}$\thanks{E-mail:
fujita@vega.ess.sci.osaka-u.ac.jp (YF)} and Hiroshi Nagai$^{2}$
\\
$^{1}$Department of Earth and Space Science, Graduate School of
 Science, Osaka University, Toyonaka, Osaka 560-0043, Japan\\
$^{2}$National Astronomical Observatory of Japan, Osawa 2-21-1, 
 Mitaka, Tokyo 181-8588, Japan
}

\date{Accepted XXX. Received YYY; in original form ZZZ}

\pubyear{2016}

\begin{document}
\label{firstpage}
\pagerange{\pageref{firstpage}--\pageref{lastpage}}
\maketitle

\begin{abstract}
We report the detection of a new feature at the centre of NGC~1275 in
the Perseus cluster, hosting the radio source 3C~84. This feature
emerges $\sim 2$~mas ($\sim 0.8$~pc) north of the central core in recent
15 and 43~GHz VLBA images, and seems to be the counterjet to a known
radio jet expanding to the south of the core. Apparently, the two jets
were born through an outburst around 2005. From the ratio of the
apparent lengths of the two jets from the core, we found that the jet
angle to the line of sight is $\theta=65^\circ\pm 16^\circ$, which is
not much different from the angle of the outer jets generated by an
activity around 1959 and constrains theories on gamma-ray emission from
jets. The new northern jet has a strongly inverted spectrum in contrast
with the southern jet. This suggests that the central black hole is
surrounded by a subparsec-scale accretion disk with the density of $\ga
10^5\rm\: cm^{-3}$. The brightness of the counterjet suggests that the
disk is highly inhomogeneous. The ambient gas density in the direction
of the jet is $\sim 8\rm\: cm^{-3}$ if the current jet activity is
similar to the past average.
\end{abstract}




\section{Introduction}

NGC~1275 is the central galaxy of the Perseus cluster. It is known as a
nearby Seyfert galaxy ($z=0.0176$) and hosts the compact radio source
3C~84. Cosmic rays accelerated around the central supermassive black
hole (SMBH) may be playing an important role in offsetting radiative
cooling of the cool core of the cluster \citep{fuj12a,fuj13a}. The
proximity of the object allows us to make detailed observations about
the environment around the SMBH and its activities. Early VLBI
observations showed that 3C~84 has complicated structures on a scale of
pc; a mushroom-like jet is expanding southward from a bright compact
core. The apparent velocity of the southern jet has been estimated to be
$\sim 0.3\: c$ \citep{1982IAUS...97..291R,asa06a,lis09a}. From this
velocity, it has been suggested that this expanding jet relates to an
outburst in 1959 \citep[e.g.][]{nes95a}. The counterjet of this southern
jet has been discovered to the north of the core
\citep{ver94a,wal94a}. The ratio of the apparent distances of the two
jets from the core suggests that the observing angle to the jet
direction is $\sim 30^\circ$--$60^\circ$ \citep{wal94a,asa06a}. The
angle between the jets and the line of sight has great importance for
the gamma-ray spectrum of NGC~1275 \citep{abd09b,ale14b,tab14a}.

Recently, new activity in the core has been reported. \citet{nag10a}
showed that a new component (C3 in the paper) emerged in the central
subparsec region of the core (C1). They indicated that this component
relates to a radio outburst that began in 2005. Moreover, since the
gamma-ray luminosity of NGC~1275 started to increase around 2005
\citep{dut14a}, the new component is seemingly associated with the
gamma-ray activity. Since the new component is moving toward south from
the core, its counterjet, if any, is expected to appear to the north of
the core.

In this Letter, we report the discovery of the northern counterjet.
From the ratio of the apparent lengths of the southern and northern jets
from the core, we estimate the inclination angle of the jets. We also
discuss the environment in the vicinity of the SMBH based on the
inverted spectrum of the northern jet. We adopt $H_0=70\rm\: km\:
s^{-1}\: Mpc^{-1}$, $\Omega_{\rm m}=0.3$, and $\Lambda=0.7$. For these
cosmological parameters, 1~mas corresponds to 0.36~pc.

\section{Data}

We use the calibrated very long base line array (VLBA) data of 3C~84
provided by Monitoring Of Jets in Active galactic nuclei with VLBA
Experiments
(MOJAVE)\footnote{\url{http://www.physics.purdue.edu/MOJAVE/sourcepages/0316+413.shtml}}
and Boston University (BU) Blazar Monitoring
Program\footnote{\url{http://www.bu.edu/blazars/VLBA_GLAST/0316.html}}.
The data were taken on 22 January 2016 at 15~GHz and on 5 December 2015
at 43~GHz. The imaging was done using CLEAN algorithm implemented in the
Difmap
software\footnote{\url{https://science.nrao.edu/facilities/vlba/docs/manuals/oss2013a/post-processing-software/difmap}}. Data
reduction was performed using the NRAO Astronomical Imaging Processing
System (AIPS)\footnote{\url{http://www.aips.nrao.edu/index.shtml}}.

\begin{figure}
\includegraphics[width=\columnwidth]{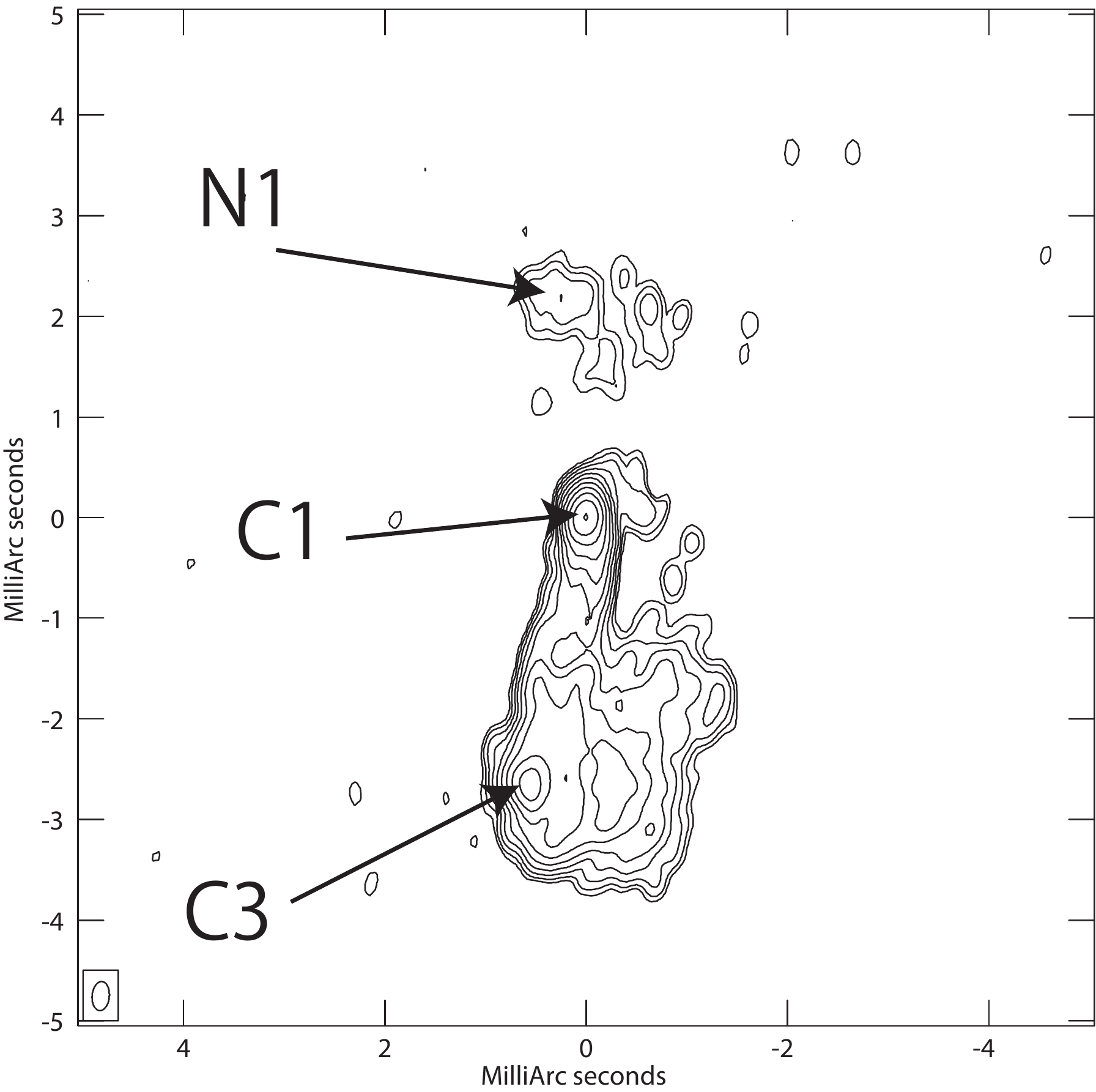} \caption{VLBA image
    of 3C~84 at 43~GHz. The core, the southern hotspot and the north
    component are indicated as C1, C3, and N1, respectively. The contour
    levels are $2.9$~mJy$\:\times\: n$ ($n=1,2,4,.....,1024$). Beam
    FWHM is $0.289\times 0.171$~mas at a position angle of
    $-4.^\circ 93$.}  \label{fig:43GHz}
\end{figure}

\begin{figure}
\includegraphics[width=\columnwidth]{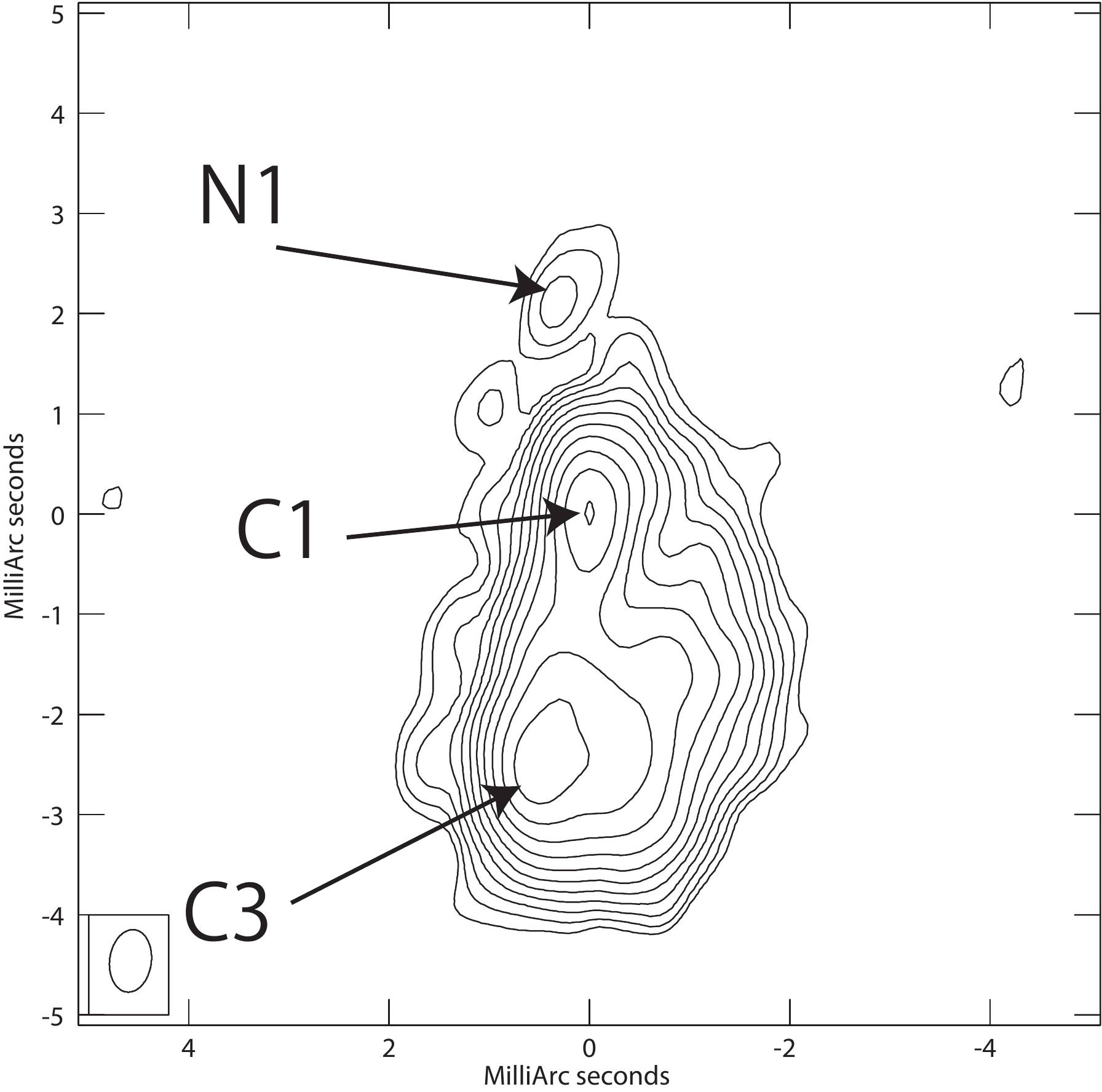} \caption{VLBA image
    of 3C~84 at 15~GHz. The core, the southern hotspot and the north
    component are indicated as C1, C3, and N1, respectively. The contour
    levels are $3.0$~mJy$\:\times\: n$ ($n=1,2,4,.....,1024$). Beam
    FWHM is $0.626\times 0.416$~mas at a position angle of
    $-7.^\circ 41$.}  \label{fig:15GHz}
\end{figure}

Fig.~\ref{fig:43GHz} shows a 43 GHz image of 3C~84.  From the core (C1),
a lobe-like feature extends southward, which we call the southern jet.
The hotspot (C3) is clearly seen in the jet. The hotspot location seems
to be shifted to the east compared to the previous 43 GHz VLBA image
\citep{2016AN....337...69N}.  A new component, which we call N1, is
recognised to the north of the core. While the background noise level is
$0.97\:\rm mJy\: beam^{-1}$, the peak flux of N1 is $23\:\rm mJy\:
beam^{-1}$, which means $24\:\sigma$ detection.  Fig.~\ref{fig:15GHz}
shows a 15 GHz image of 3C~84.  The north component N1 can also be seen
at the same position in Fig.~\ref{fig:43GHz}. The background noise level
is $1.0\:\rm mJy\: beam^{-1}$ and the peak flux of N1 is $17\:\rm mJy\:
beam^{-1}$, which means $17\:\sigma$ detection.  Thus, it is unlikely
that the detection is false.

\section{Angle to the Line of Sight}
\label{sec:angle}

From the ratio of apparent lengths of the southern and north jets from
the core and from the apparent jet velocity, we can estimate the angle
of the jets to the line of sight $\theta$. The ratio $D$ is represented
by
\begin{equation}
\label{eq:D}
 D = \frac{1+\beta\cos\theta}{1-\beta\cos\theta}\:,
\end{equation}
where $\beta$ is the jet velocity normalised by the light speed $c$
\citep{wal94a}. The observed velocity of the approaching jet is given by
\begin{equation}
\label{eq:Ba}
 \beta_{\rm a} = \frac{\beta\sin\theta}{1-\beta\cos\theta}
\end{equation}
\citep{wal94a}. Thus, the inclination angle is
\begin{equation}
\label{eq:theta}
 \theta = \arctan\frac{2\beta_a}{D-1}
\end{equation}
\citep{asa06a}. \citet{nag10a} estimated that the apparent velocity of
the approaching component (C3) is $\beta_{\rm a}=0.23\pm 0.01$.

We use the 43~GHz image in Fig.~\ref{fig:43GHz} to derive the ratio $D$,
because the resolution is higher than the 15~GHz image in
Fig.~\ref{fig:15GHz}.  We assume that the tips of the southern and
northern jets are C3 and N1, respectively. The position of C3 and N1 are
derived by fitting them with a gaussian using AIPS task {\sf JMFIT}. We
ignore the error of the position associated with the fit, because it is
much smaller than the beam size ($0.289\times 0.171$~mas). The position
of the core (C1) is derived in the same way, and we assume that it is
the position of the black hole. The ratio $D$ is given by the ratio of
the distance from the black hole to C3 to the distance from the black
hole to N1, and it is $D=1.22\pm 0.16$, where the error comes from the
beam size. Thus, we obtain $\theta=65^\circ\pm 16^\circ$ from
equation~(\ref{eq:theta}), and $\beta=0.23\pm 0.02$ from
equations~(\ref{eq:D}) or (\ref{eq:Ba}). The obtained angle is not much
different from that of the outer pair of 'old' jets ($\theta\sim
30^\circ$--$60^\circ$) associated with an outburst of 3C~84 around 1959
\citep{wal94a,asa06a}. Moreover, the jet direction on the sky is similar
to that of the old jets \citep{ver94a,wal94a}. These suggest that the
jet direction has not much changed over the last half a century. 
Note that the position of the core may be affected by an opacity-induced
core shift. For M~87, the core shift is 0.007--0.01~pc at 43~GHz
\citep{had11a}, which corresponds to $\sim 0.02$~mas for NGC~1275. Thus,
the effect of the core shift is probably negligible. However, Fig.~6 of
\citet{wal00a} shows that the core has a strong northern ridge for the
inner mas or so at higher frequencies. The position of the peak of the
core may depend on resolution due to blending of the inner
features. Thus, as an extreme case, we set the position of the black
hole at the northern edge of the core (0.3~mas north of the peak), and
we calculate the jet parameters. We found that $D=1.56\pm 0.21$,
$\theta=39^\circ\pm 10^\circ$, and $\beta=0.28\pm 0.04$. The angle
$\theta$ is still fairly large.

We expect that the 'new' jets (C3 and N1) are associated with gamma-ray
emission that started to increase around 2005 \citep{abd09b,dut14a}. The
obtained inclination angle has an implication for theories on gamma-ray
radiation from jets in radio galaxies. A popular idea is that NGC~1275
is a misaligned blazar. However, although one-zone synchrotron-self
Compton models may explain the observed broadband spectral energy
distribution (SED), they require an untypical low bulk-Lorentz factor
\citep{ale14b}. Even if two-zone 'spine-layer' jet models are
considered, they conflict with the relatively large angle that we
obtained ($\theta=65^\circ\pm 16^\circ$). This is because the
gamma-rays with energies above a few tens of GeV are absorbed in the
luminous infrared radiation field associated with the fast spine
emission \citep{tab14a}. Moreover, in spite of the large inclination
angle, the gamma-ray luminosity of NGC~1275 is very large and is
comparable to those of blazars \citep{kat10a}. These facts suggest that
the misaligned blazar interpretation may not be applied for NGC~1275.

\section{Absorption by the accretion disk}

The spectral index of the radio emission can give us information on
absorption toward an object. Thus, we estimate the spectral indices for
the southern and the northern components. For that purpose, we make an
image at 43~GHz with a convolving beam of $0.626\times 0.416$~mas to
match the resolution of the 15~GHz observation
(Fig.~\ref{fig:15GHz}). Then, we estimate the fluxes of the components
C3 and N1 by fitting them with a gaussian using AIPS task {\sf JMFIT}
for both of the 43 and 15~GHz images. From the ratio of the flux at
43~GHz to that at 15~GHz, we find that the spectral index of C3 is
$\alpha_{\rm S}=-0.91$ and that of N1 is $\alpha_{\rm N}=1.61$, where
the spectral index is defined as $S\propto \nu^\alpha$. The index of C3
is the one for a typical synchrotron emission region, which means that
the component is not much affected by absorption. On the other hand, N1
has an inverted spectrum and it is likely to be absorbed. Considering
the inclination angle of the jets ($\theta\sim 65^\circ$), a natural
explanation is that the black hole is surrounded by an optically-thick
accretion disk, which is perpendicular to the jets and obscures only the
northern component.

If the southern and the northern components are symmetric, and if there
is no absorption, the ratio of brightness between the two components (C3
and N1) is written as
\begin{equation}
 R=\left(\frac{1+\beta\cos\theta}{1-\beta\cos\theta}\right)^{\alpha+m}
\;,
\end{equation}
where $m$ is ether 2 or 3 for a continuous jet or single component,
respectively \citep{wal94a}. Using the values of $\theta$ and $\beta$
derived in section~\ref{sec:angle} and assuming that $\alpha=\alpha_S$,
we can predict that the brightness ratio should be $R=1.8\pm 0.6$ for
$m=2$, and $R=2.1\pm 1.0$ for $m=3$. However, the observed ratios are
much larger and they are $R_{{\rm obs}}=45$ at 43~GHz and $R_{\rm
obs}=600$ at 15~GHz. These large values clearly indicate that the north
component is actually absorbed by the accretion disk.

The most likely explanation for the inverted spectrum and the large
ratio of brightness is free-free absorption \citep{lev95a}. The
free-free optical depth can be written as
\begin{equation}
\label{eq:tau}
 \tau_{\rm ff} \approx 5.6\times 10^{-8}\bar{g}
\left(\frac{T}{10^4\rm K}\right)^{-3/2}
\left(\frac{n_{\rm e}}{\rm cm^{-3}}\right)^2
\left(\frac{\nu}{\rm GHz}\right)^{-2}
\left(\frac{L}{\rm pc}\right)\:,
\end{equation}
where $T$ is the temperature of the absorbing medium, $n_{\rm e}$ is the
electron density \citep[e.g.][]{ryb79a}, and $L$ is the disk depth along
the line of sight. Here, we assumed that the absorbing medium is a pure
hydrogen plasma with a uniform density, and that the electron density
equals the proton density. For $T\sim 10^4$~K and $\nu\sim 15$--43~GHz,
the Gaunt factor is $\bar{g}\sim 4$ \citep[e.g.][]{hum88a}. We can
estimate the optical depth from the relation of
\begin{equation}
 R_{\rm obs} = R\exp(\tau_{\rm ff}) \:.
\end{equation}
For $m=2$, the optical depth is $\tau_{\rm ff}\approx 3.2$ at 43~GHz,
and $\tau_{\rm ff}\approx 5.8$ at 15~GHz. We can calculate the density
of the disk using equation~(\ref{eq:tau}). Assuming that $T=10^4$~K and
$L=0.8$~pc, which is the apparent distance between C1 and N1, the
density is $n_{\rm e}\approx 1.9\times 10^5\rm\: cm^{-3}$ for 43~GHz,
and $n_{\rm e}\approx 8.9\times 10^4\rm\: cm^{-3}$ for 15~GHz. The
density is not much different when $m=3$. If the disk is thinner or $L$
is smaller, the density is higher. Thus, we conclude that the gas
density is $n_{\rm e}\ga 10^5\rm\: cm^{-3}$.

It is interesting to note that the ratio of the observed optical depths
at 43 and 15~GHz suggests that $\tau_{\rm ff}\propto \nu^{-0.6}$, which
is different from $\tau_{\rm ff}\propto \nu^{-2}$ in
equation~(\ref{eq:tau}). One explanation may be that the absorbing
medium is highly inhomogeneous and it consists of regions of $\tau_{\rm
ff}\gg 1$ and $\tau_{\rm ff}\ll 1$. If this is the case, the averaged
optical depth is less dependent on the observational frequency.

If a temperature of $T=10^4$~K is assumed, the optical depth $\tau_{\rm
ff}$ we obtained and equation~(\ref{eq:tau}) mean that the emission
measure (EM) is $n_{\rm e}^2 L\sim 2.7\times 10^{10}\rm\: pc\: cm^{-6}$
for 43~GHz and $\sim 6.2\times 10^9\rm\: pc\: cm^{-6}$ for 15~GHz at the
position of N1 (a projected distance of $d=0.8$~pc from the core). On
the other hand, \citet{wal00a} found that the EM at $d\sim 1.4$--3.6~pc
is $n_{\rm e}^2 L \sim 5.7\times 10^8\: (d/{2.7\rm\: pc})^{-1.5}\rm\:
pc\: cm^{-6}$ based on the data taken in January 1995 or $n_{\rm e}^2 L
\sim 4.8\times 10^8\: (d/{2.7\rm\: pc})^{-1.8}\rm\: pc\: cm^{-6}$ based
on the data taken in October 1995\footnote{\citet{wal00a} adopted
$H_0=75\rm\: km\: s^{-1}\: Mpc^{-1}$ and we have corrected the
difference from ours.}. If we extrapolate these to $d=0.8$~pc, the EM is
$n_{\rm e}^2 L \sim 4\times 10^9\rm\: pc\: cm^{-6}$. Thus, the EM we
obtained is $\sim 1.5$--7 times larger than the extrapolated
one. However, this may be because \citet{wal00a} derived their profiles
for a fairly narrow range of distance. For $d\sim 3$--30~pc,
\citet{sil98a} indicated that the EM is $n_{\rm e}^2 L \sim 1\times
10^6\: (d/30\rm\: pc)^{-2.6}\rm\: pc\: cm^{-6}$ if $T=10^4$~K. If we fit
the above data over the whole range ($d\sim 0.8$--30~pc) with a single
power-law, they are well represented by $n_{\rm e}^2 L \sim 7\times
10^9\: (d/\rm pc)^{-2.6}\rm\: pc\: cm^{-6}$.


\section{Gas density in the jet direction}

\citet{fuj16a} discussed the density of the surrounding medium in the
direction of the outer old jets or perpendicular to the accretion disk
in NGC~1275. Here, we apply this argument to the new jets shown in
Figs.~\ref{fig:43GHz} and~\ref{fig:15GHz}.

The momentum balance along the jet is given by
\begin{equation}
\label{eq:Lj}
 L_{\rm j}/c = \rho_{\rm h}V_{\rm h}^2 A_{\rm h}\:,
\end{equation}
where $L_{\rm j}$ is the kinematic power of the jet, $\rho_{\rm h}$
is the density of the ambient gas just ahead of the jet, $V_{\rm h}$ is
the velocity the hotspot (the jet head), and $A_{\rm h}$ is the
cross-section area of the jet head. The density profile of the ambient
gas is assumed to be
\begin{equation}
 \rho(r) = \rho_{\rm h}(r/r_{\rm h})^{-X}\:,
\end{equation}
where $r$ is the three-dimensional distance from the black hole and
$r_{\rm h}$ is the position of the jet head. Note that since the new
jets are propagating in the old jets, the ambient gas here means the gas
in the old jets. We implicitly assume that the old jets are filled with
a significant fraction of thermal gas. The age of the jet is given by
\begin{equation}
\label{eq:tage}
 t_{\rm age} = \frac{2\: r_{\rm h}}{4-X}
\left(\frac{L_{\rm j}}{\rho_{\rm h} c A_{\rm h}}\right)^{-1/2}
\end{equation}
from equation~(8) in \citet{fuj16a}\footnote{In \citet{fuj16a}, the
density profile is given by $\rho(r)=\rho_{\rm B}(r/r_{\rm B})^{-X}$,
where $r_{\rm B}$ is the Bondi radius and $\rho_{\rm B}=\rho(r_{\rm
B})$. Thus, equation~(\ref{eq:tage}) in this letter can be obtained
using the relation of $\rho_{\rm h}r_{\rm h}^X=\rho_{\rm B}r_{\rm
B}^X$.}.  

The apparent distance between the black hole and C3 is $\sim
2.7$~mas. Considering the inclination angle ($\theta\sim 65^\circ$), the
actual jet length is $r_{\rm h}\sim 1.1$~pc. The width of the southern
lobe at the position of C3 is $\sim 1.9$~mas (Fig~\ref{fig:43GHz}), and
thus the area of the jet head is $A_{\rm h}\sim 0.36\rm\: pc^2$. The jet
velocity is $V_{\rm h}\sim 0.23\: c$
(section~\ref{sec:angle}). Following \citet{fuj16a}, we assume $X=1.5$,
although results are insensitive to the value of $X$. We also assume
that the current jet power $L_{\rm j}$ is comparable to the average jet
power in the past ($\sim 10^7$~yr). From the size of X-ray cavities,
\citet{raf06a} estimated that the average power is $P_{\rm cav}\sim
1.5\times 10^{44}\rm\: erg\: s^{-1}$.

Based on these assumptions, we can derive $\rho_{\rm h}$ so that the
condition of $2\: L_{\rm j}=P_{\rm cav}$ is satisfied using
equation~(\ref{eq:Lj}). We find that the number density of the ambient
gas at $r=r_{\rm h}$ is $n_{\rm e}(r_{\rm h})\sim 8.1\rm\:
cm^{-3}$. The age of the jets is $t_{\rm age}=12$~yr
(equation~(\ref{eq:tage})), which means that the jets was born in 2003
or 2004. The birth time is not much different from the beginning of the
recent activity around 2005 \citep{abd09b,dut14a}. Note that for the
outer old jets, \citet{fuj16a} obtained $n_{\rm e}(r_{\rm h})\sim
0.096\rm\: cm^{-3}$ for $2\: L_{\rm j}=0.37\: P_{\rm cav}$ and $r_{\rm
h}=5.6$~pc. If $2\: L_{\rm j}=P_{\rm cav}$, the density is $n_{\rm
e}(r_{\rm h})\sim 0.26\rm\: cm^{-3}$.

\section{Time variability}

It is interesting to note that there was a feature at the position
of N1 in the full resolution 15 and 22 GHz images shown in Fig.~5 of
\citet{wal00a} and in the deep 23 GHz image shown in Fig.~1 of
\citet{wal03a}. The observations were made in October 1995 for the
former and in September 1998 for the latter. Although 15~GHz data
(MOJAVE) have been available since 1999, noticeable features are not
recognised at the position of N1 before September 2015. For 43~GHz data
(BU), which have been available since 2010, a noticeable feature appears
at the position of N1 in December 2015. Since both MOJAVE and BU
observations are not deep, we cannot deny that dim features had been
overlooked.

If there has always been a feature there, it might be the result of some
other effect. For example, there may be a groove on the accretion disk,
and when a bright component in the jet (e.g. a knot) passes behind the
groove, N1 may be brightened. If the emergence of N1 at the end of 2015
is attributed to the passage of the hot spot of the new jet behind the
groove, the main conclusions of this paper do not change. On the other
hand, N1 may be associated with an internal structure in the jet that
stays at the same position (e.g. a re-collimation shock) while the
luminosity changes. Long and deep observations are required to
investigate these possibilities.

\section{Conclusions}

We discovered a new feature $\sim 2$~mas ($\sim 0.8$~pc) north of the
central core of NGC~1275 (3C~84) at 15 and 43~GHz with VLBA. This
feature is considered to be the counterjet of the jet expanding
southward from the core that launched around 2005. From the ratio of the
lengths of the two jets, we estimated the inclination angle of the jet
and found that it is $\theta=65^\circ\pm 16^\circ$, which is not much
different from that of the outer old jets that launched around 1959. The
northern jet has a strongly inverted spectrum, which indicates that it
is absorbed by an accretion disk around the SMBH. From the brightness of
the northern jet, we calculated the density of the disk and found that
it is $\ga 10^5\rm\: cm^{-3}$. We also indicated that the disk may be
highly inhomogeneous. Assuming that the current jet power is not much
different from the past average, we derived the ambient gas density in
the jet direction ($\sim 8\rm\: cm^{-3}$).

\section*{Acknowledgements}

We would like to thank the referee for useful comments. We are grateful
to Motoki Kino, and Kiyoaki Wajima for helpful discussion. This work was
supported by MEXT KAKENHI No.~15K05080 (YF). HN is supported by MEXT
KAKENHI No.~15K17619. This research has made use of data from the MOJAVE
database that is maintained by the MOJAVE team
\citep{2009AJ....137.3718L}. This study makes use of 43~GHz VLBA data
from the VLBA-BU Blazar Monitoring Program (VLBA-BU-BLAZAR;
http://www.bu.edu/blazars/VLBAproject.html), funded by NASA through the
Fermi Guest Investigator Program. The VLBA is an instrument of the
National Radio Astronomy Observatory. The National Radio Astronomy
Observatory is a facility of the National Science Foundation operated by
Associated Universities, Inc.








\bsp	
\label{lastpage}
\end{document}